\documentclass[10pt]{iopart}

\usepackage{hyperref}

\usepackage{iopams}

\usepackage{graphicx}
\graphicspath{ {images/} }
\usepackage{subcaption} 

\usepackage{subfiles}
\usepackage{standalone}
\usepackage{pdfpages}
\usepackage[toc,page]{appendix} 
\usepackage{float} 

\usepackage[font=footnotesize,labelfont=bf]{caption}
\usepackage{lipsum}
\usepackage{flushend}
\usepackage{booktabs}
\usepackage{multicol}

\makeatletter
\def\footnoterule{\kern-3\p@
  \hrule \@width 2in \kern 2.6\p@} 
\makeatother

\begin{document}

\title[]{\center Genetic algorithm with cross validation-based epidemic model and application to early diffusion of COVID-19 in Algeria}

\author{M. T. Rouabah\footnote[1]{{Corresponding author:} 
m.t.rouabah@gmail.com, 
rouabah.taha@umc.edu.dz.},
 A. Tounsi, N. E. Belaloui }

\address{Laboratoire de Physique Math\'{e}matique et Subatomique\\
Fr\`{e}res Mentouri University Constantine - 1, Ain El Bey Road, Constantine, 25017, Algeria.}

\vspace{10pt}

\begin{abstract}

A dynamical epidemic model optimized using  genetic algorithm and  cross validation method to overcome the overfitting problem is proposed. 
The cross validation procedure is applied  so that available data are split into a training subset used to fit the algorithm’s parameters, and a smaller subset used for validation. 
This process is tested on the countries of Italy, Spain, Germany and South Korea before being applied to Algeria. 
Interestingly, our study reveals an inverse relationship between the size of the training sample and the number of generations required in the genetic algorithm. 
Moreover, the enhanced compartmental model presented in this work is proven to be a reliable tool to estimate key epidemic parameters and  non-measurable asymptomatic infected portion of the susceptible population in order to establish realistic nowcast and forecast of epidemic's evolution. 
The model is employed to study the COVID-19 outbreak dynamics in Algeria between February 25th and May 24th, 2020. The basic reproduction number and effective reproduction number on May 24th, after three months of the outbreak, are estimated to be 3.78 (95\% CI 3.033-4.53)  and 0.651 (95\% CI 0.539-0.761) respectively. Disease incidence, CFR and IFR are also calculated. 
Numerical programs developed for the purpose of this study  are made publicly accessible for reproduction and further use.
\end{abstract}


\ioptwocol

\section{Introduction}
The outbreak of the highly infectious COVID-19 disease attributable to SARS-CoV-2 in Wuhan and other provinces in China in 2019 has become a global pandemic since the first quarter of 2020 as asserted by the World Health Organization (WHO). 
Epidemic modeling, along with biological and medical research \cite{Su2020,Rothan2020}, might significantly contribute in understanding the outbreak’s epidemic characteristics. It is deemed to be a crucial tool in predicting the inflection point and ending time and provides insights into the epidemiological situation. Such analysis can predict the potential future evolution, help estimate the efficiency of already taken measures, and guide the design of alternative interventions \cite{Ajelli2010,Chowell2019,Atangana2020,Gao2020}.
The compartmental classical Susceptible Exposed Infectious Recovered (SEIR) model \cite{Anderson92,Meloni2011} has been the most widely adopted model for characterizing many historical propagating infectious diseases such as the Spanish flu \cite{Chowell2007}. SEIR model is extensively used to study the COVID-19 pandemic in China and many other countries with variations best suiting the subject region and time period \cite{Wu2020,Peng2020,Labadin2020}.
Since finding the correct parameters of such a dynamic model for an epidemic is essentially a curve fitting problem, the predictive effectiveness of the model can considerably be reduced if  the available real data, which we will call training data, are  underfited or overfited. Indeed, if the training data are underfitted, the model could simply diverge or give overestimated numbers with very large variance. On the opposite, if the data are overfitted, the predictive curves produced by the model will be strongly influenced by the given training data and will have very low variance, thus, artificially reducing the error on the predicted numbers and eventually leading to a non-realistic forecast. Overfitting remains a major problem with epidemic dynamical models \cite{Yang2020}. 
In many of them, overfitting 
arises as a consequence of the fact that multiple parameters might fluctuate over their uncertainty ranges causing their fitted values to be extremely susceptible to noise in the original data \cite{Basu2013}. Therefore, restrictions have been applied to some epidemic analysis including COVID-19 outbreak in order to diminish the number of free parameters and inhibit overfitting, affecting the pertinence of those studies \cite{Peng2020}. To overcome this issue, our approach in this study consists of using a genetic fitting algorithm and cutting off the fitting process after a number of generations which is large enough to actually fit the training data and small enough to not go beyond overfitting limits. We will call this number the optimum generation number $G_{opt}$ and will compute it using data from a given province or country passing through a two-samples cross validation procedure. Thus, $G_{opt}$ corresponds to the fitting depth that ensures the well balance between underfitting and overfitting in our model.

SARS-CoV-2 was first imported to Algeria on Feb. 17th, 2020 by an Italian national who has been confirmed positive to COVID-19 on Feb. 25th \cite{AHM-1}. The Italian man has been repatriated via a special flight on Feb. 28th and no contaminated individuals by this first confirmed case have been reported by the Algerian official authorities \cite{AHM-2}. As far as we know, the effective outbreak of COVID-19 in Algeria started late Feb. 2020. Indeed, the Algerian Health Ministry (AHM) reported in a statement on March 2nd the two first confirmed cases of COVID-19 in Blida province south of the capital Algiers \cite{AHM-3}. Since then the spread of the virus in Algeria has gone through different epidemic phases \cite{Lounis2020}. 
To the best of our knowledge, few theoretical studies on COVID-19 outbreak in Algeria have so far been achieved \cite{Bentout2020,Belkacem2020}. The lack of theoretical and clinical publicly accessible studies about SARS-CoV-2 spread in Algeria exposing the actual situation and analyzing possible evolution scenarios is making the situation more confusing for the Algerian public and scientific community. A lot of studies on COVID-19 specifications and dynamics around the world are published every day, some of which include the Algerian case \cite{Zhao2020,Luo2020}.
However, we believe that any analysis of COVID-19 outbreak in Algeria should take into consideration many specific aspects that are not considered in such universal studies and online-simulators which use raw data accessible on many databases. Beyond the fact that the majority of those databases contain many wrong reported data for Algeria, data nomenclature and interpretation, as well as test methods proper to every country should be taken into consideration for more accurate outcomes. 
In our analysis instead of relying only on official Reverse Transcriptase Polymerase Chain Reaction (RT-PCR) confirmed SARS-CoV-2 infection cases, which are strongly affected by limited test capacities, we rather combine them with the official number of hospital admitted patients due to SARS-CoV-2 infection in order to deduce the effective number of new confirmed infection per day. This choice makes a significant difference not only on the cumulative number of confirmed cases but also on the nowcast and forecast of the virus spread.

The paper is organized as follows: in the next section we present the mathematical model we use for the dynamical modeling of COVID-19 propagation and some results of the model with reference to the pandemic spread in Italy,  Spain,  Germany  and  South  Korea. The third section will be devoted to the application of the model on the Algerian case through the estimation of key epidemic parameters and a forecast analysis. Results will be shown and discussed in the fourth section. The concluding section will include some ideas about future developments of this work.

\section{Model and Methods}

At the very beginning of the epidemic, during the free spread phase, 
initial exponential-growth is commonly assumed, which is distinctive of most human infectious diseases \cite{Anderson92,Yuan2020}. However, spontaneous herd immunity, protections and lockdown measures will confront the geometrical evolution. A dynamical model is then required to describe the evolution of the disease.
\subsection{Compartmental SEIQRDP model}

Regarding the novelty of the time course of infection shown by the disease and the required protection measures, to simulate COVID-19 spread we use a SEIQRDP model in which at time $t$ the population is split
between different compartments representing different stages in the course of the disease \cite{Peng2020,Tang2020}. 
The susceptible portion of the population i.e. individuals yet to be infected, is represented by the compartment $S(t)$.
$P\left(t\right)$ represents the effectively protected population, mainly individuals who tend to strictly follow the standard advised protection measures such as wearing masks, physical distancing ... etc. Hence, this part of the population is considered as not susceptible to be infected. Introducing this compartment is crucial to reflect increasing awareness within the major part of the population as the pandemic evolves and allows to take into consideration the control measures taken by authorities to fight against the pandemic such as closing public areas, suspending public transportation and lockdown. $E(t)$ 
covers individuals that have been exposed to the virus but are not infectious yet. This compartment represents a latent state in which individuals are infected but cannot infect other individuals.
whereas $I(t)$ 
represents individuals that are currently infectious.
The asymptomatic exposed and infectious portions of the population are not detectable and hence non-measurable. The proportion of this part of the population can only be revealed by theoretical modeling of the disease. $Q(t)$ represents quarantined individuals considered as active cases, $R(t)$ 
corresponds to the portion of population that has recovered from the disease and supposed to be no longer involved in the virus propagation 
and $D(t)$ represents closed cases or deaths. $N=S(t)+E(t)+I(t)+Q(t)+R(t)+D(t)+P(t)$ is the total population at time $t$ considered constant at the time scale of the epidemic evolution. The SEIQRDP model represents the virus propagation by a 
collection of ordinary differential equations associating a set of transition parameters to the movement of individuals among the population compartments defined above:
\begin{eqnarray}
\dot{S} &= -\beta S (t) I(t) / N - \alpha S(t)\\
\dot{E} &= \beta S(t)I(t)/N-\gamma E(t)\\
\dot{I} &= \gamma E(t)-\delta I(t)\\
\dot{Q} &= \delta I(t) - \lambda Q(t) - \kappa Q(t) \\
\dot{R} &= \lambda Q(t) \\
\dot{D} &= \kappa Q(t) \\
\dot{P} &= \alpha S(t)
\end{eqnarray} 
where $\dot{S}$ refers to the time derivative of $S$. The positive rate $\alpha $ called the protection rate, is introduced into the model assuming that the susceptible population is steadily decreasing as a result of increasing population awareness and public health authorities’ actions \cite{Peng2020}. All the other parameters depend on the evolution of the epidemic, testing and health care capacities, and are calculated based on the official daily confirmed cases, deaths and recoveries numbers. The transmission rate $\beta $ represents the ability of an infected individual infecting others (depending on the population density, the toxicity of the virus etc \ldots) and $\beta S\left(t\right)I\left(t\right)/N$ is the incidence of the disease, i.e., the number of new infected individuals yielding in unit time at time $t$ \cite{Ma2009}. $\gamma ^{-1}$ is the average latent time that an individual spends incubating the virus to become infectious (infected but not yet infectious) and $\delta ^{-1}$ is the average infectiousness time, i.e, time for an infectious individual to get symptoms and get detected and quarantined. $\lambda $ is the cure rate and $\kappa $ is the mortality rate while $\lambda ^{-1}$ and $\kappa ^{-1}$ represent respectively quarantine to recovery time and quarantine to death time. These transition parameters are used by the model to define a time-dependent number of secondary cases generated by a primary infectious individual so-called the effective reproduction number  $\mathcal{R}_{t}=\beta \delta ^{-1}S\left(t\right)/N$. It is a very important parameter for the analysis of any epidemic outbreak and yields a measure of
severity of interventions necessary to overcome the virus spread. 
In general, if $\mathcal{R}_t >1$, which mathematically corresponds to $\dot{E}+\dot{I}>0$, the disease propagates epidemically and when $\mathcal{R}_{t}<1$, the disease is vanishing. At the beginning of the epidemic matching a situation of a fully susceptible population, this quantity is known as the basic reproduction number $\mathcal{R}_{0}=\beta \delta ^{-1}$ and is obtained by the next generation matrix method \cite{Diekmann2013}.

Even though many COVID-19 studies try to calculate universal mean values of the reproduction number and transition parameters in some specific spots of the outbreak, they remain strongly related to local data and could change from one country to another and even from one region to another within the same country. Such parameters are the kind of valuable information this model could provide in addition to approximate peak times of the disease (infection peak time and active cases peak time) and approximate numbers of the non-measurable asymptomatic cases, active cases, and total quarantined, recoveries and deaths cases. An a priori knowledge of those numbers, though approximate, could help to optimize human and material resources on the global and local scales of a country. In this SEIQRDP model, key parameters are extracted from official numbers of cumulative confirmed cases, recoveries and deaths available at a given period of the epidemic. The parameters obtained either by direct calculation or by a fitting algorithm are used to construct the variables curves that fit the initial data. Those curves are then extrapolated to a longer period, thus forecasting the evolution of the epidemic.

\subsection{Fitting with a genetic algorithm}

To calibrate our model’s parameters and fit the real data originating from a specific region of the world, many fitting methods are available, most of which are widely used in epidemiological studies and machine learning models. Single stage problems such as calibrating the parameters of our model are usually solved with modified deterministic optimization methods such as the L-BFGS-B method \cite{Hannah2015,Comunian2020}. However, a stochastic method would have the benefit of taking into account the diversity of the possible calibrations’ scenarios. Evolutionary genetic algorithms are one of those stochastic methods that have a good reputation in solving optimization problems. In the following section we will discuss the advantages that the genetic algorithm method can yield to our study.

\subsection*{Definition}
A genetic algorithm (GA) is an optimization approach inspired by Darwinian evolution in which an initial set of candidate solutions called initial population, each represented by a set of values making a genome, evolve by breeding and reproducing while being subject to random mutations \cite{Chambers95}. The key mechanism in a GA is that only the best performing solutions get to reproduce and pass on their genes just as in Darwinian natural selection. The evolution process is finally stopped after a certain number of generations when a defined stopping condition is met.

\subsection*{Application}
In our case, applying a GA to find the best fitting for the SEIQRD model parameters is a straightforward application of the above definition; The genome is the set of parameters itself and the breeding is the process of creating a new set from two parent sets by randomly selecting genes from either one of the parents, and mutation is a random alteration of one of the genes of the resulting new genome. The best performing set of parameters is the one for which the curves produced by the model match the best the original data. This is measured by a normalized least squares method. We can speed up the process by constraining the randomly generated initial population to be somewhat around already published values for COVID-19 epidemic parameters \cite{Lin2020}. Different runs of the GA give slightly different solutions. From these solutions, an error on the prediction made by the model can be computed.

\subsection*{Computation of the optimum generation number using cross validation}
Cross validation is a procedure where an original data set is split into training and validation subsets, and where the model is trained on the first subset and tested for the second one \cite{Browne2000}. In our case, the original data set is the whole available data on the COVID-19 epidemic for a given country or region for $n$ days. Those data are then split into a training subset containing the data of the first $n-v$ days, and a validation subset of the last $v$ days. The ratio $v/n$ depends on the number of adjustable parameters in the regression problem \cite{Guyon97}. This ratio is around $1/4$ for the SEIQRDP model.

To determine the optimum generations number $G_{opt}$,
corresponding to the fitting depth that ensures avoiding both underfitting and overfitting in our model, we run the genetic algorithm for fitting with the training subset and measure after each generation the fitness of the best solution with the validation set. The expected result of this process is a bad fitness for very low generations number $G$, which gets better with every new generation until we start overfitting the training subset (too large values of $G$) resulting in a worse fitness. The value of $G\equiv G_{opt}$ for which the fitness on the validation set is the best is chosen as the stopping point for the genetic algorithm when applied for predictive purposes.

\begin{figure*}[t!]
\centering
\begin{subfigure}[b]{0.475\textwidth} 
    \hspace{-0.5cm}
    \includegraphics[width=\textwidth]{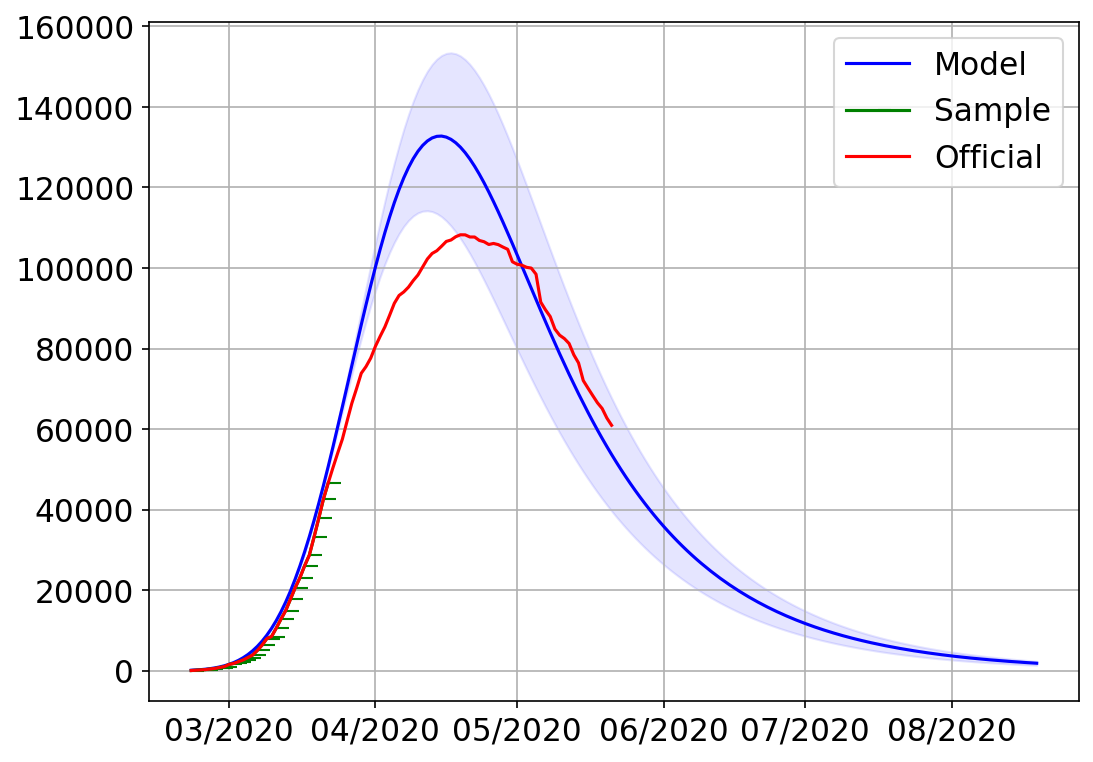}
    \caption{Italy, $G_{\rm opt} = 40, S_N = 17\%$.}    
    \label{fig:Italy}
    \end{subfigure}
\begin{subfigure}[b]{0.475\textwidth} 
    \centering 
    \includegraphics[width=\textwidth]{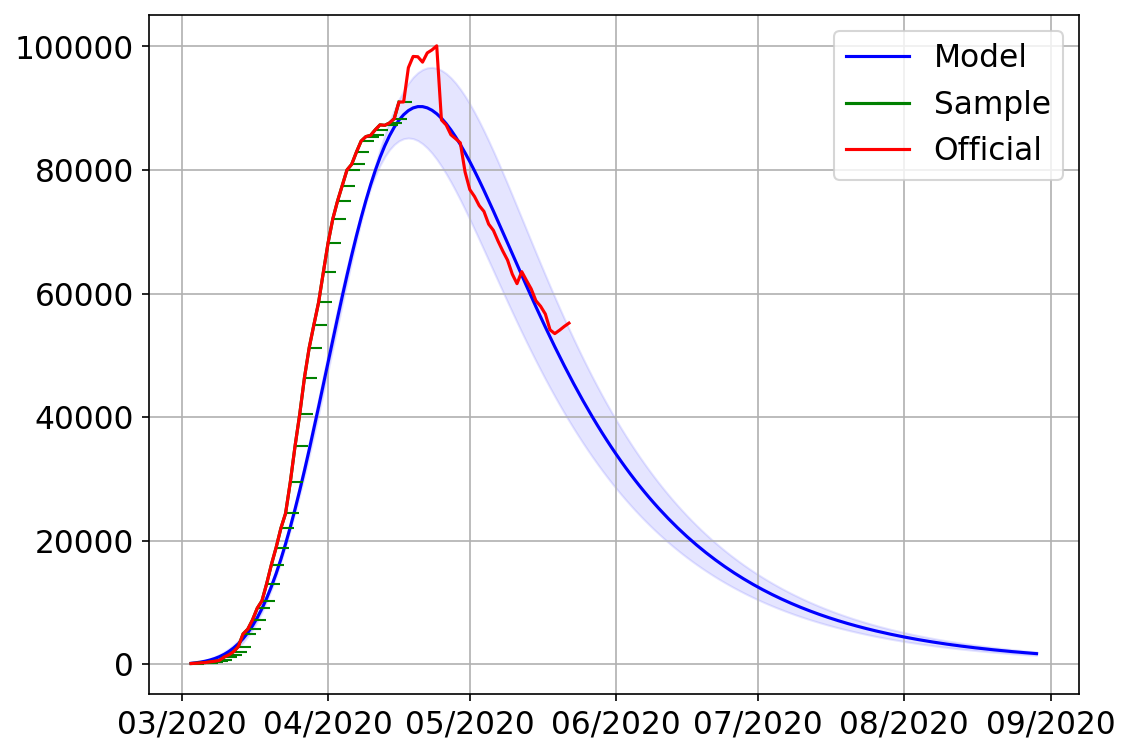}
    \caption{Spain, $G_{\rm opt} = 20, S_N = 10\%$.}  
    \label{fig:Spain}
    \end{subfigure}
\begin{subfigure}[b]{0.475\textwidth}
    \centering
    \includegraphics[width=\textwidth]{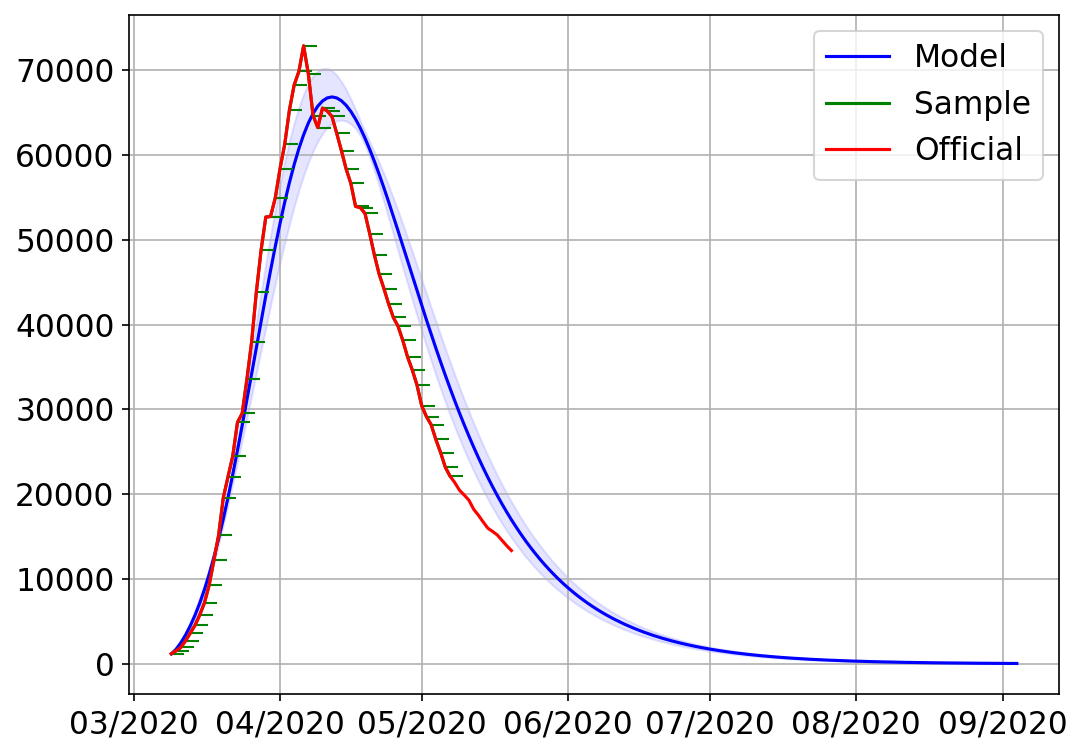}
    \caption{Germany, $G_{\rm opt} = 10$, $S_N = 7\%$.}    
    \label{fig:Germany}
    \end{subfigure}
    \quad
\begin{subfigure}[b]{0.475\textwidth}   
    \centering 
    \includegraphics[width=\textwidth]{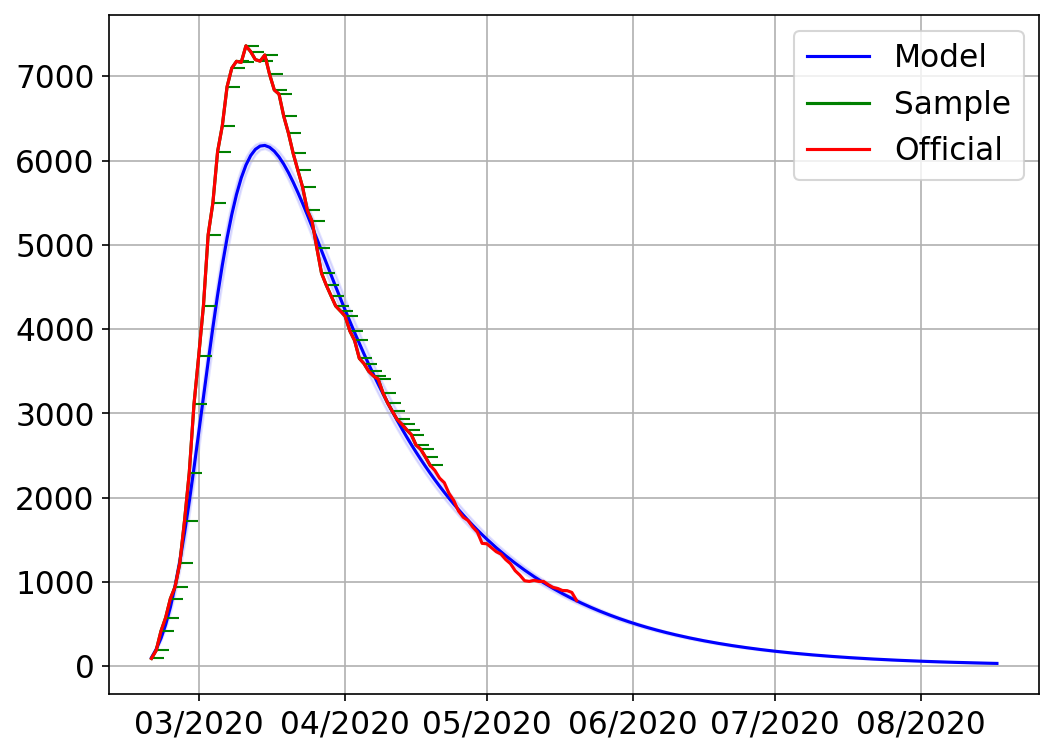}
    \caption{South Korea, $G_{\rm opt} = 10, S_N = 3\%$.}    
    \label{fig:SK}
    \end{subfigure}
        \caption{Results of the SEIQRDP model forecast for COVID-19 outbreak in (a) Italy, (b) Spain, (c) Germany and (d) South-Korea. A training sets of 30, 45, 60 and 90 days of official data (green tri\_down marker lines), respectively, are initially used to fit the model's parameters. The forecast curves (blue lines) are calculated using the SEIQRDP model and compared to real active cases curves (red lines) for each country. Light blue shading represents 95\% confidence intervals of the model estimate. Even with only four or six weeks of  training data the model is able to produce a realistic forecasting estimate. All fits have coefficient of determination $R^2 > 0.9$. 
        }
        \label{fig: world-wide simulations}
\end{figure*}

\subsection{Computational tools}
In order to allow the readers to take advantage of the fitting algorithm and the cross validation method presented in this study for other epidemic cases, a tailored set of Python programs developed by the authors have been gathered in a Python package and made accessible online with all the necessary instructions for installation and efficient use \cite{GitHub-repo}. This package is adapted for parallel computation and includes tools to: download data from online repositories, determine the optimum fitting depth for a given city, region or country using the cross validation method, calibrate the SEIQRDP model by fitting the real data with the genetic algorithm, and solve the system of ODEs to produce the forecast. Hence, our generic programs might be easily applied to study any outbreak for which a compartmental analysis is adequate in any region of the world provided that a sufficient amount of epidemic data is available. 

\begin{figure*}[t!]
    \centering
    \includegraphics[width=\textwidth]{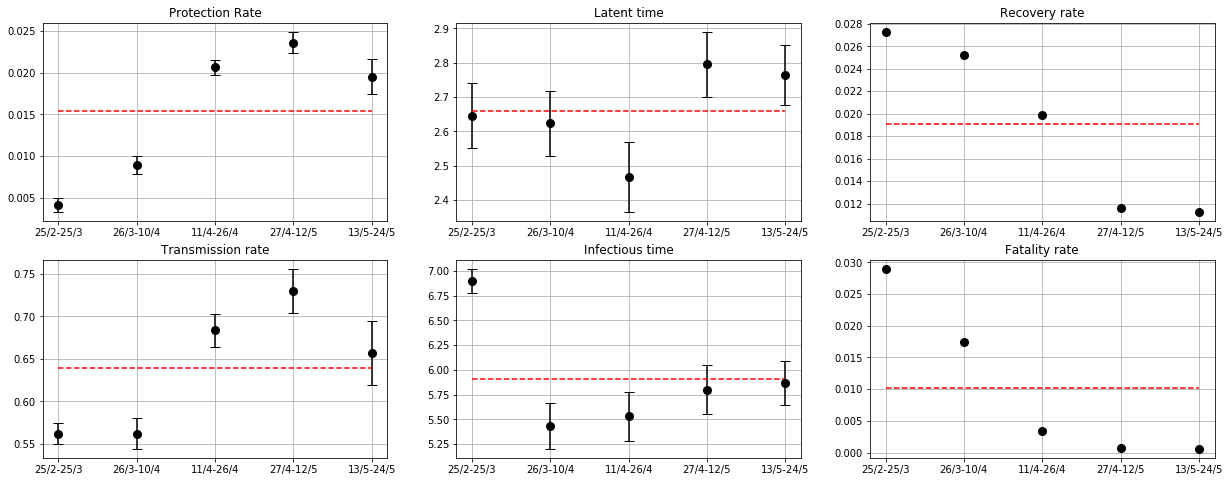}
    \caption{Estimation of key epidemic parameters  during  the early stage of COVID-19 outbreak in Algeria (Feb. 25th - May 24th, 2020). Intermediate values are calculated for five different time periods corresponding to specific phases of the virus propagation  with specific circumstances and authorities measures severity. The intermediate values of each parameter are compared to its mean value  on the whole three months period (dashed red line). Protection rate, transmission rate, latent time and infectious time are estimated using the SEIQRDP model while recovery and fatality rates are calculated from official data. Error bars represent 95\% confidence intervals of the model estimate.}
    \label{fig:seiqrdp_parameters}
\end{figure*}

\subsection{Model validation}
Provided reasonably accurate data, our model successfully reproduces the evolution of COVID-19 in different spots worldwide for which a sufficient amount of data is available. In this section we present the results obtained using the enhanced SEIQRDP model to estimate the active cases evolution in Italy, Spain, Germany and South Korea. For those countries we use publicly available confirmed cases, recoveries and deaths numbers from online raw data sets \cite{Covid-data,Ritchie-data}. We pick training data starting from the date for which all confirmed cases, recoveries and deaths numbers take non-zero values to avoid computational bugs and optimize parameters fitting. The active cases curve is then reproduced for 6 months following that date.
In order to highlight the efficiency of the model we use only an early part of the official data to train the model instead of all the available data. We have used 30 days training data for Italy, 45 days for Spain, 60 days for Germany and 90 days for South Korea. Fig.\ref{fig:Infected} shows that the results obtained using our model are in very good accordance with official statistics for the number of active cases in those countries. All the fittings have a coefficient of determination $R^{2}>0.9$ and as our fitting method is based on a non-linear regression algorithm, we use a normalized standard error $S_{N}$ of the estimate to evaluate the goodness of the fits. We remind that in order to avoid overfitting the training data and obtain the best forecast, we look for the optimum fitting corresponding to $G_{opt}$ rather than the best one. For Italy, the model is able to reproduce to a good approximation $\left(S_{N}=17\% \right)$ the active cases curve with only 30 days of training data.
Fig.\ref{fig: world-wide simulations} reveals that the larger is the training data sample the lower is the optimum number of generations $G_{opt}$ used by the genetic algorithm to fit the data with a lower $S_{N}$. The active cases curve is one of the most pertinent in our opinion as it reflects the amplitude of the epidemic outbreak as well as the efficiency of the measures applied to control it. Moreover, the epidemic will end only if all the active cases are closed. Germany and South Korea are very special cases and need profound analysis that is beyond the scope of this paper, but one can clearly observe the quicker decrease in their active cases after the epidemic peak time due to particular strategies to control the epidemic.

\section{Model estimation for Algeria}

\subsection{Data}
For COVID-19 dynamics study in Algeria, we use official public data provided by the Algerian Health Ministry (AHM) \cite{AHM-EADN,CDTA-map}. Our analysis specificity, that we believe makes its predictive results for Algeria more accurate than different studies in which Algeria is presented as an example \cite{Zhao2020}, is the fact that instead of relying on official numbers of RT-PCR-confirmed SARS-CoV-2 infection cases, which are strongly affected by limited test capacities, we deduce the effective number of confirmed infections per day by considering the number of hospital admitted patients.
This number--unfortunately, no longer provided by the AHM after May 25th, 2020--is considered as the effective number of active cases in our study. The effective confirmed cases number for a given date is then deduced by adding computed tomography (CT) scans confirmed cases to the official RT-PCR confirmed cases (see \ref{apx: A}).

\begin{figure*}[t!]
\centering
\begin{subfigure}[b]{0.475\textwidth} 
    \centering 
    \includegraphics[width=\textwidth]{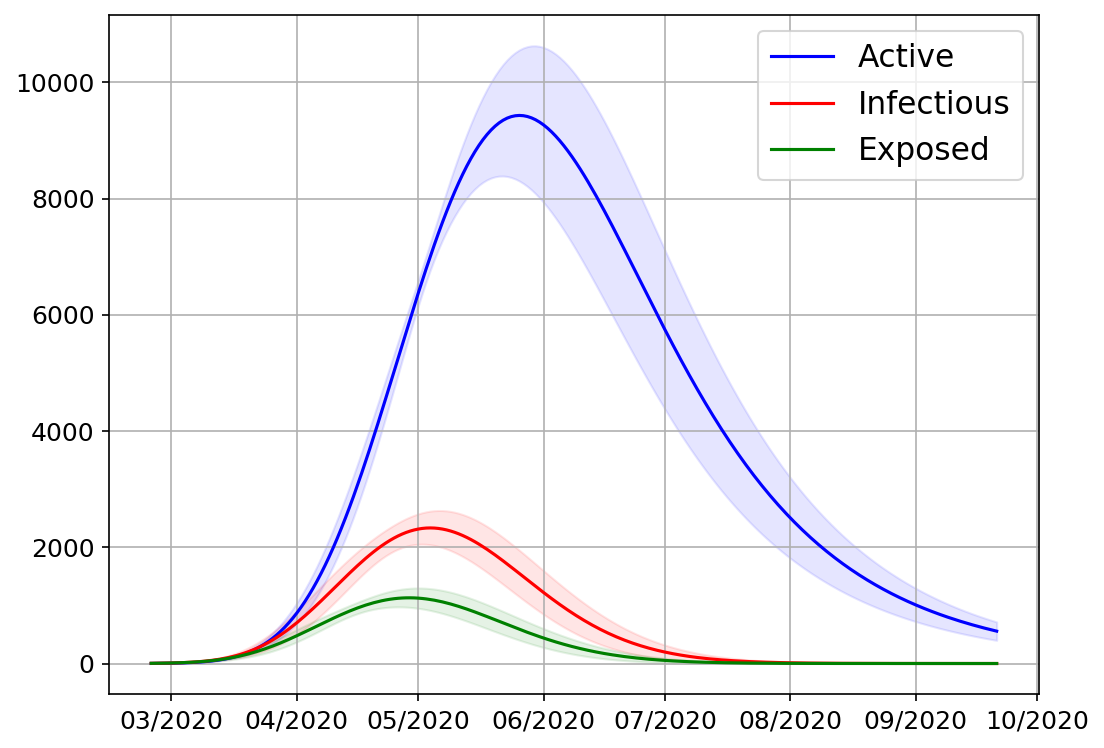}
    \caption{Infected Cases.}    
    \label{fig:Infected}
    \end{subfigure}
\hfill
\begin{subfigure}[b]{0.475\textwidth} 
    \hspace{-0.5cm}
    \includegraphics[width=\textwidth]{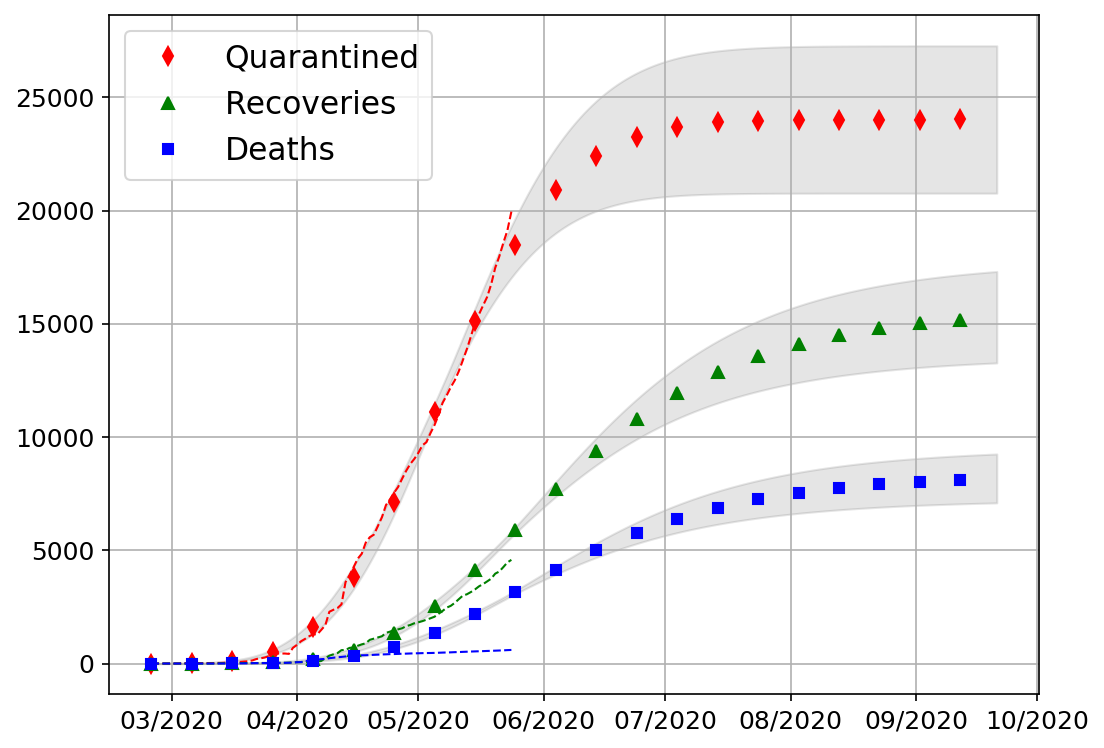}
    \caption[]{Cumulative numbers.}    
    \label{fig:Cumulative}
    \end{subfigure}
\begin{subfigure}[b]{0.475\textwidth}
    \centering
    \includegraphics[width=\textwidth]{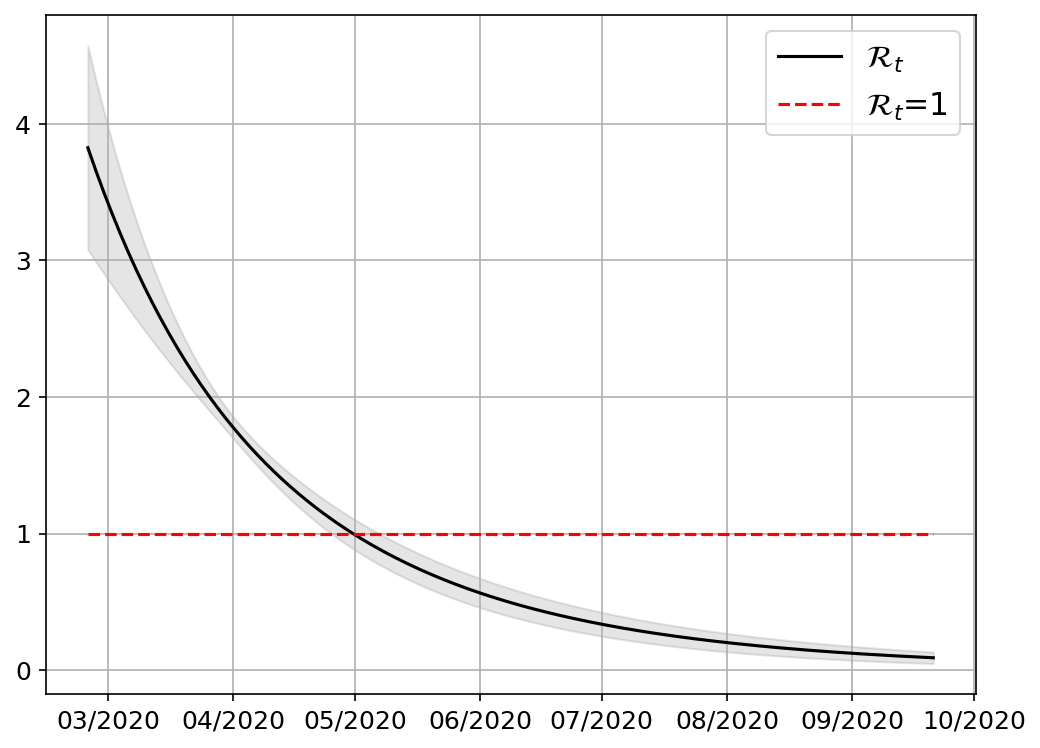}
    \caption[]%
            {Reproduction number since Feb. 25th, 2020.}    
    \label{fig:Reproduction Number}
    \end{subfigure}
    \quad
\begin{subfigure}[b]{0.475\textwidth} 
    \centering 
    \includegraphics[width=\textwidth]{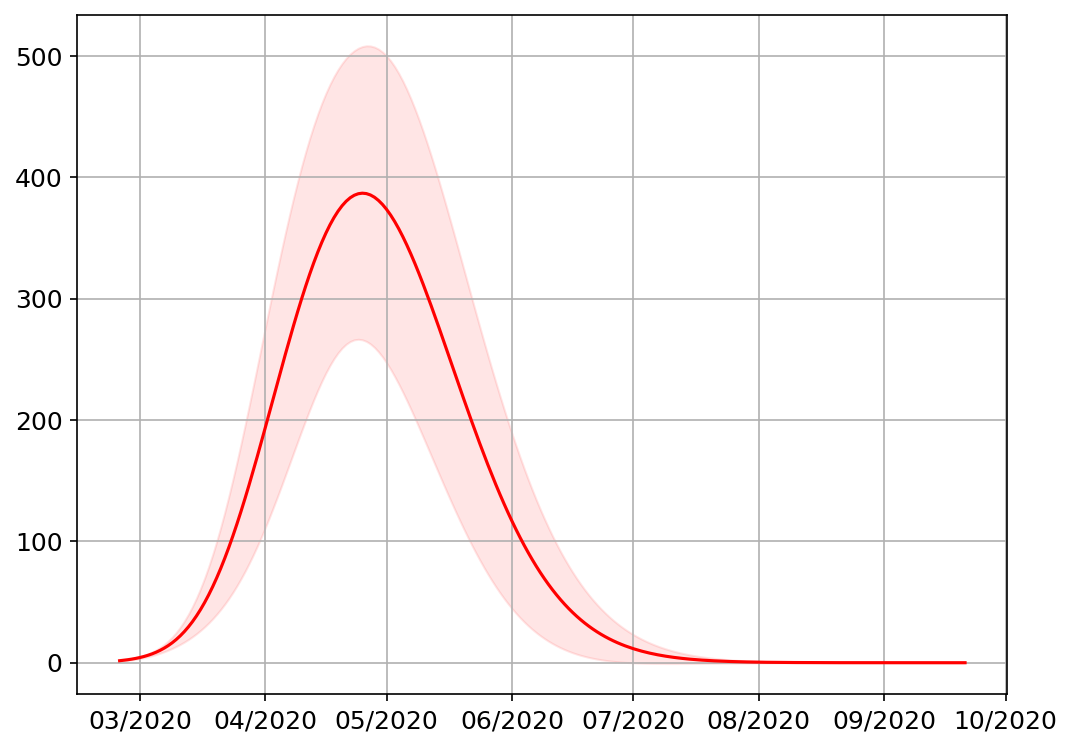}
    \caption[]{Incidence of the disease.}  
    \label{fig:Incidence}
    \end{subfigure}
        \caption{SEIQRDP model forecast. (a) Number of exposed (infected not yet infectious), infectious (asymptomatic infectious) and active (quarantined) cases. The figure shows the epidemic peak time corresponding to the maximum active cases to be on the time period of May 20th - May 30th with roughly ten thousand active cases. (b) Total quarantined, recoveries and deaths. Real data are represented with the red, green and blue dashed lines respectively. (c) Time dependent reproduction number.  $\mathcal{R}_t=1$ point is in a perfect accordance with exposed, infected and active cases inflection points. (d) Number of new infected individuals per day. Light shadings represent 95\% confidence intervals of the model estimate.}
        \label{fig:forecast}
\end{figure*}
 
\subsection{Epidemic parameters}
Besides the more exciting forecasting use of the SEIRQDP model, this latter is particularly efficient for nowcasting. Indeed, fitting the official data allows us to estimate key epidemic parameters of the early stage spread of COVID-19 in Algeria. Even though hundreds of studies are estimating those parameters for COVID-19 in different spots of the world, a local estimation is of major importance as their values are strongly related to local population discipline, public health capacities and severity of local containment measures at the very beginning and during the epidemic period.

During one month after the first confirmed case of COVID-19 in Algeria on Feb. 25th, 2020 the disease has undergone a practically free propagation phase. On March 12th, universities, schools and nurseries were closed. On March 19th, all trips between Algeria and European countries have been canceled by Algerian authorities who have decided the first strong containment measures against COVID-19 spread on March 24th. A total lockdown of Blida province and partial lockdown in many other provinces have been applied. Coffee shops, restaurants and all non-essential shops have been closed, public transportation suspended and grouping of more than two persons forbidden. On April 24th, the authorities decided a partial release of lockdown measures in Blida and other provinces and allowed many commercial activities to resume. This date coincided with the starting of the holy month of Ramadan resulting in a brutal increase of social and commercial activities. Due to low respect of physical distancing and protection measures the number of new confirmed cases increased significantly and shops have been closed again in many provinces since May 7th. 
In the light of this chronology of measures we have estimated intermediate mean values of the epidemic parameters during the free-propagation phase (Feb. 25th - Mar. 25th) and then every two-weeks intermediate period till May 24th. Those intermediate mean values exposed in Fig.\ref{fig:seiqrdp_parameters} provide valuable information revealing the evolution of the epidemic in Algeria during its three first months and the impact of the applied control measures. 

\begin{table*}[t!]
\centering
\begin{tabular*}{\textwidth}{@{}l@{\extracolsep{\fill}}llll@{}}
\toprule
\begin{tabular}[c]{@{}l@{}}Parameter\\ \ \end{tabular}      & \begin{tabular}[c]{@{}l@{}}Definition\\ \ \end{tabular}                & \begin{tabular}[c]{@{}l@{}}Value for Algeria\\ (95\% CI)\end{tabular}     & \begin{tabular}[c]{@{}l@{}}Value for Wuhan\\ (95\% CI) \end{tabular} & \begin{tabular}[c]{@{}l@{}}Reference\\ (Wuhan)\end{tabular}                                       \\ \midrule
\ $\alpha$  & Protection rate (mean)  & 0.015  (0.014-0.017)        & 0.085                                                               & Peng et al. \cite{Peng2020}\\
\ $\beta$  & Transmission rate  (mean)  & 0.64  (0.62-0.66)  & 0.99     & Peng et al. \cite{Peng2020}\\
\ $\gamma^{-1}$  & Latent time      (mean)   & 2.7 (2.6-2.8)  & 2  & Peng et al. \cite{Peng2020} \\
\ $\delta^{-1}$ & Infectious time (mean)  & 5.9  (5.7-6.1)  & 7.4    & Peng et al. \cite{Peng2020}  \\
\begin{tabular}[c]{@{}l@{}}\ $\mathcal{R}_0$ \end{tabular} & \begin{tabular}[c]{@{}l@{}}Basic reproduction number \end{tabular} & \begin{tabular}[c]{@{}l@{}}3.78 (3.033-4.53) \end{tabular} & \begin{tabular}[c]{@{}l@{}} 6.47 (5.71–7.23)\end{tabular}     & \begin{tabular}[c]{@{}l@{}} Tang et al. \cite{Tang2020}
\end{tabular} \\ \bottomrule
\end{tabular*}
\caption{Summary of SEIQRDP parameters estimates for early stage of COVID-19 outbreak in Algeria compared to Wuhan (China). $\mathcal{R}_0$ is estimated on Feb. 25th while the mean values of the other parameters are calculated for the three first months of the outbreak.}
\label{table}
\end{table*}

\subsection{Forecast}

In order to forecast the evolution of the COVID-19 in Algeria we apply the SEIRQDP with a training data period from Feb. 25th to May 24th. The cross validation method script is applied on the first 70 days of the data set and tested on the 20 remaining to calculate the optimum number of generations. For the chosen data sample, we obtain $G_{opt}=20$. Then, the genetic algorithm and the rest of SEIQRDP set of programs are applied on the whole training data to calculate the optimum fit and reproduce the SEIQRDP variables curves using the fit parameters obtained. We present in this paper a forecast of COVID-19 outbreak dynamics until the end of September 2020, time for which the reopening of schools and universities is scheduled. 
That step might represent a turning point in the disease's epidemic evolution and requires a specific analysis.

\subsection{Results and Discussion}
Our model estimates that on Feb. 25th, in addition to the first confirmed SARS-CoV-2 infected case in Algeria at least 7 other individuals have been infected without showing any symptoms. On March 2nd when the two first cases have been confirmed at Blida, we estimate that the number of asymptomatic infectious people has already reached 10 individuals and at least 10 others have been in a latent period. One week later the number of asymptomatic infected persons have already exceeded 70 following our estimations. Officially, 20 of them have been confirmed at that time.

Epidemic parameters model estimates for the first three months of COVID-19 in Algeria are in a good agreement with on-the-ground evolution of the outbreak. The estimated basic reproduction number on Feb. 25th is $\mathcal{R}_{0}=3.78$ (95\% CI 3.033-4.53) while the value of $\mathcal{R}_{t}$ on May 24th is estimated to 0.651 (95\% CI 0.539-0.761) and the mean effective reproduction number during the first three months of the epidemic is evaluated to 1.74 (95\% CI 1.55-1.92). The notable decline in $\mathcal{R}_{t}$ during this period might reflect outbreak control efforts and 
 increasing consciousness of COVID-19.
By the same token, we distinguish a significant rise of the protection rate ${\alpha}$ after the first control measures on March 24th jumping from 0.0041 during the free propagation phase before March 25th to 0.0089 on the period of March 26th - April 10th and doubling again to 0.021 on the next period (see Fig.\ref{fig:seiqrdp_parameters} upper-left corner). Interestingly, the protection rate curve reflects the release of containment and a lower respect of protection measures in the period between April 27th and May 12th resulting in a decline of ${\alpha}$ during the next period. The protection rate mean value of the overall study period is estimated to 0.015 (95\% CI 0.014-0.017). 
The increase of the transmission rate shown on Fig.\ref{fig:seiqrdp_parameters} lower-left corner is reasonable due to the continuous propagation of the virus and the apparition of many clusters in dense population provinces. In addition, the low number of daily tests and the relatively long test-to-result time of the used testing technology increase the probability that an asymptotic infectious individual spread the virus before being quarantined. The transmission rate mean value is estimated to 0.64 (95\% CI 0.62-0.66). The mean latent time is evaluated to 2.7 (95\% CI 2.6-2.8) days and the mean infectious time is predicted to 5.9 (95\% CI 5.7-6.1) days. 
The mean incubation time (latent time + infectiousness time) has a mean value of 8.6 (95\% CI 8.3-8.9) days. One remarkable point that can be observed on Fig.\ref{fig:seiqrdp_parameters} middle panel is that besides the first period, the incubation time remains relatively stable taking values within the range [7.9-8.6] days.
This reflects the fact that the model effectively calibrates the global features of the evolution of hidden variables representing the exposed $E\left(t\right)$ and infectious $I\left(t\right)$ portions of the population which are not measurable. 
The decrease of the incubation time after the first period of the study might be a consequence of a better detection scheme. In fact, a high diagnosis capacity allowing large scale testing strategy and efficient tracking are essential tools to reduce the onset to quarantine (incubation) period since early and quick detection of infectious individuals enables authorities to quarantine them before showing symptoms, hence limiting the number of their contacts. Moreover, this will help diminish the effective reproduction number $\mathcal{R}_{t}$ and then better control the disease spread.

In contrast to other epidemic parameters, recovery and fatality rates shown on the right panel of Fig.\ref{fig:seiqrdp_parameters} are directly calculated from official data. The recovery rate varies in the range [1.1\% - 2.7\%] with a mean value of 1.9\% and the fatality rate, initially estimated as the highest in the world at the time, fell below 0.5\% since mid-April with a mean value estimated to 1.02\%. The significant decrease of fatality rate, even though affected by the growing test capacities after the number of RT-PCR daily tests have been increased and the CT-scan diagnostic of COVID-19 adopted in the beginning of April, could also be interpreted as the consequence of better medical care. The fatality rate seems to stabilize during the last month of the study (0.071\% on April 27th - May 12th and 0.058\% on May 13th - May 24th) as newly deployed RT-PCR test capacities are reaching again their limits. The epidemic analysis of the parameters’ values is beyond the scope of this paper as it requires information to which we don’t have access. Nevertheless, we notice that the key parameters values obtained through our model for Algeria fall within the values ranges estimated for the Chinese city of Wuhan where SARS-CoV-2 first appeared \cite{Peng2020,Tang2020,Lin2020} as shown in Table \ref{table}.

The forecast simulations (Fig.\ref{fig:forecast}), based on the available official data, estimate that the infection peak time corresponding to the maximum incidence occurred on April 24th-26th with 387 (95\% CI 267-509) new infections per day as shown on Fig.\ref{fig:Incidence}. The effective reproduction number continuously decreased reflecting a better control of the disease spread and crossed the line $\mathcal{R}_{t}=1$ by May 1st (see Fig.\ref{fig:Reproduction Number}). At that crucial point the disease entered the attenuation phase. The SEIQRDP model evaluates the active cases peak time for the first wave of COVID-19 outbreak in Algeria, corresponding to active cases maximum, to be on the period between May 20th and May 30th with 9794 (95\% CI 8770-1024) active cases (see Fig.\ref{fig:Infected}).

We estimate that the number of new infections will vanish by mid-September. At that time the number of active quarantined cases will be still above 500. Assuming that the epidemic will remain ongoing as long as all active cases have not been closed yet, the model predicts the outbreak's first  wave to end no earlier than October 2020, with an estimated total quarantined individuals of 24021 (95\% CI 20768-27274), 15291 (95\% CI 13272-17310) recovered and 8172 (95\% CI 7093-9251) deaths as shown on Fig.\ref{fig:Cumulative}. Notice that the predicted total number of deaths appears to be particularly overestimated compared to official numbers (blue dashed line). A solution to this technical issue is under investigation \cite{Nikolaou2020}. This can partly be explained by the fact that official COVID-19 deaths are those confirmed by PCR tests only while our model deals with both PCR and CT scan diagnosed individuals.
We emphasize that the numbers we present in this forecast are only estimations that could be seriously affected by the quality of the available data.

\begin{figure}[t!]
    \centering
    \includegraphics[width=\columnwidth]{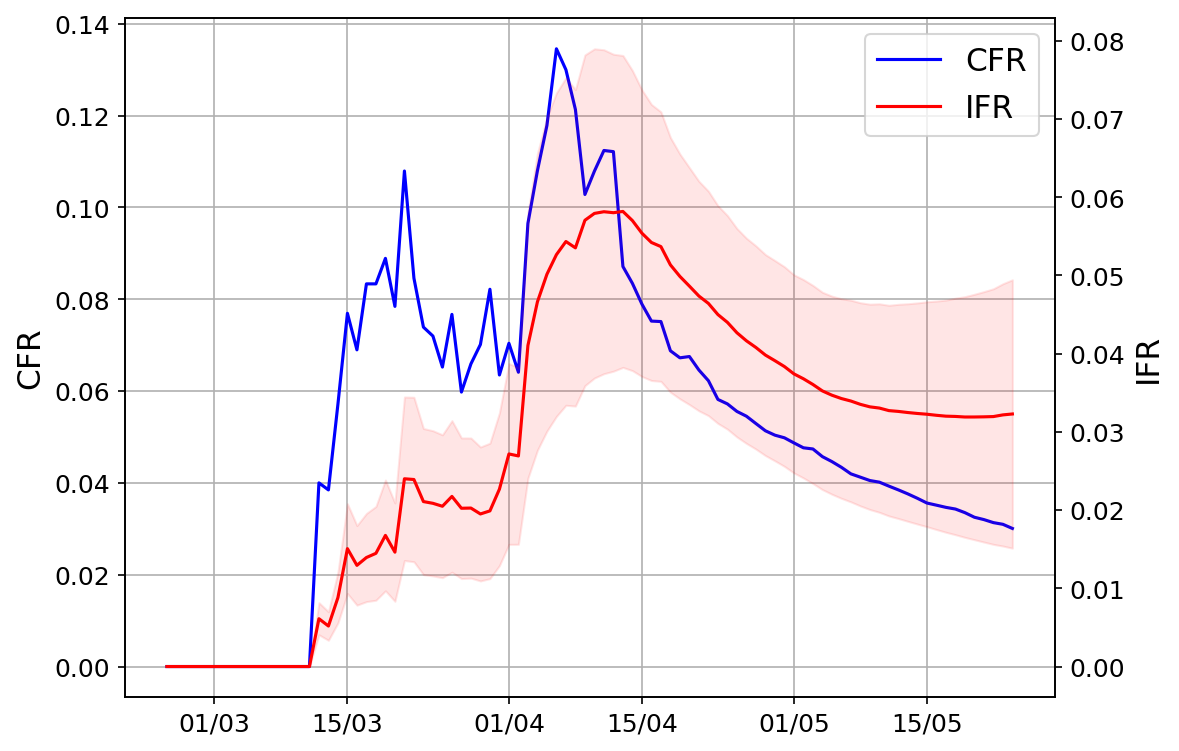}
    \caption{Case to Fatality Rate (CFR) and Infection to Fatality Rate (IFR) for COVID-19 outbreak in Algeria between Feb. 25th and May 24th, 2020. Light red shading represents 95\% confidence intervals of the model estimate.}
    \label{fig:CFR-IFR}
\end{figure}

Another important piece of information that could be extracted from the official public data is the Case Fatality Rate (CFR) corresponding to the ratio of deaths to effective confirmed cases.
Furthermore, the Infected Fatality Rate (IFR), often confused with CFR, is the ratio of deaths to infected cases including asymptomatic cases which are non-measurable. For that reason, we calculate CFR based on official data while the IFR is calculated through the ratio of the official cumulative deaths to the cumulative number of infected individuals obtained from the SEIQRDP model (see Fig.\ref{fig:CFR-IFR}). The mean CFR on the period Feb. 25th - May 24th is estimated to be 5.3\% while the mean value of IFR on the same period is 2.9\% (95\% CI 1.7\%-3.9\%). Notice that the mean IFR for the three first months of the outbreak in Algeria is higher than the global value estimated to 1.4\% by a recent study using cumulative COVID-19 data from 139 countries \cite{Grewelle2020}.

It is worth to know that compartmental models including the SEIQRDP model work perfectly when some conditions on the studied population are assumed. Indeed, the SEIQRDP model requires a well-mixed and homogeneous population. Well-mixed population means that all individuals in the population have the same chance to be infected by an infectious one. Homogeneity means that all individuals behave likely toward the disease and thus are governed by the same rules of transitions’ probabilities between different population compartments. Consequently, all calibrated parameters in this study should be seen as a statistical average over population. Moreover, SEIQRDP model is fundamentally not additive i.e the sum of different SEIQRDP models applied to different provinces of a given country is not necessarily equivalent to the SEIQRDP model applied to the whole country. Because of the previous considerations altogether, it would be very interesting to apply our study on different major infected cities of the country separately. 

\section{Conclusion}
In this paper we have presented an enhanced compartmental SEIQRDP model for epidemics in which a protection rate has been introduced and noteworthy compartments of quarantined and protected population have been added compared to the most widely used SEIR models. Our approach is based on a genetic fitting algorithm and makes use of a cross validation method to overcome the overfitting problem. 
In order to determine the optimum number of generations for the genetic algorithm yielding the best parameters, the cross validation procedure was applied on various countries in such a way that the whole available data on the COVID-19 epidemic for a given country for $n$ days is split into a training subset containing the data for the first $n-v$ days, and a validation subset for the last $v$ days. The ratio $v/n$ depends on the number of adjustable parameters in the regression problem and turns out to be around $1/4$ for our model. The procedure was tested on the countries of Italy, Spain, Germany and South Korea before applying it to Algeria. Remarkably, our study emphasis that there is an inverse relationship between the size of the training  sample and the number of generations required in the genetic algorithm. As more data becomes available for a given country, the optimum number of generations decreases. Therefore optimization should be re-adjusted at different points in outbreak period to ensure the most accurate results are obtained using the model. We have designed a generic open-source package containing all computational tools used in our analysis \cite{GitHub-repo}. This package includes tools to pick up online data, calculate the optimum fitting depth using the cross validation method, fit the real data with the genetic algorithm to estimate the SEIQRDP parameters, and produce forecast curves. Moreover, the package includes parallel computation functionality allowing for the programs to be deployed on high performance computers for better accuracy. We have neatly prepared this package in order to be easily utilized to study any epidemic for which a compartmental analysis is adequate in any regions of the world. 

Based on official cumulative recoveries, cumulative deaths and deduced effective cumulative confirmed cases--including approximated CT-scan diagnosed cases--this model allowed us to estimate epidemic parameters for COVID-19 outbreak in Algeria (basic reproduction number, protection rate, transmission rate, infectious time, latent time {\ldots}). We have estimated intermediate mean values of key epidemic parameters between Feb. 25th and May 24th. These intermediate values exposed in Fig.\ref{fig:seiqrdp_parameters} permit to evaluate the epidemic situation in the country and the effect of the different phases of control measures during the first three months of the outbreak. We recapitulated in Table \ref{table} the basic reproduction number estimated on Feb. 25th and the calculated mean values of key epidemic parameters on the whole period Feb. 25th - May 24th and compared them to recently published results for COVID-19 epidemic in Wuhan. Such parameters estimations might be of high interest for further epidemic studies of the virus spread in Algeria and the African continent. Using our SEIQRDP model, we have made a prediction of the disease effective reproduction number time evolution ($\mathcal{R}_{t}$) which is considered as an essential indicator of the epidemic situation. Fig.\ref{fig:Reproduction Number} exposes the evolution of $\mathcal{R}_{t}$ since the beginning of the outbreak and an approximate period on which this parameter has gone below one. Our simulations suggest the basic reproduction number on Feb. 25th, 2020 to be $\mathcal{R}_{0}$ = 3.78 (95\% CI 3.033-4.53) while the value of $\mathcal{R}_{t}$ on May 24th is estimated to 0.651 (95\% CI 0.539-0.761) and the mean effective reproduction number during the first three months of the epidemic is evaluated to 1.74 (95\% CI 1.55-1.92). Moreover, we have been able to provide a valuable approximate estimation of the daily evolution of the non-measurable asymptomatic exposed and infectious cases in addition to the daily active cases from the beginning until an advanced stage of the COVID-19 outbreak in Algeria (Fig.\ref{fig:Infected}, Fig.\ref{fig:Cumulative}). We have estimated the periods in which these numbers will be at their highest peak and approximated the maximum values they could reach for the outbreak's first stage. The model predicts this wave of COVID-19 epidemic in Algeria to end not sooner than October 2020, with an estimated total quarantined individuals of 24021 (95\% CI 20768-27274), 15291 (95\% CI 13272-17310) recovered and 8172 (95\% CI 7093-9251) deaths. We have also estimated the time in which the number of new infections will eventually vanish (Fig.\ref{fig:Incidence}). Even though the SEIQRDP model we presented, as many of SEIR derivatives, is effective in different contexts, we are still studying Algerian case carefully because the reported COVID-19 epidemic evolution in Algeria quickly reached the country’s maximum capacity of diagnosis which is well reflected in the linear form of official confirmed cases data. 
Furthermore, we should note that, in a basic way, the SEIQRDP model is well established to simulate outbreaks in a well-mixed closed population  although being very sensitive to data accuracy. In this instance, we stress the fact that our estimations depend strongly on the public available data at the time this study has been achieved and we emphasize the specificity of our study considering an effective cumulative confirmed cases number including approximated CT-scan diagnosed SARS-CoV-2 infections deduced from the official number of hospital admitted patients as illustrated in \ref{apx: A}. 

We are investigating many possibilities to further optimize our model to fit the COVID-19 evolution in Algeria and elsewhere with more ingenious methods. Additionally, a completely different epidemic agent-based model is already in an advanced development stage and will be used to tackle the virus spread from a different perspective. A comparison of updated epidemic parameters in Algeria using more recent data to recent studies performed on African countries \cite{amouzouvi2021} might also be of great interest.

We hope this study can serve as a useful guideline for scientists and governments and efficiently contribute to the fight against COVID-19 pandemic on national and international scale.

\section*{CRediT authorship contribution statement}
\textbf{M. T. R.}: Conceptualization, Methodology, Software, Formal analysis, Writing - Original Draft, Supervision. \textbf{A. T.} : Conceptualization, Methodology, Software, Formal analysis, Writing - Review \& Editing. \textbf{N. E. B.}: Conceptualization, Methodology, Software, Formal analysis, Writing - Review \& Editing.

\section*{Conflict of interest}
All authors declare no conflicts of interests.

\section*{Acknowledgment}
This study presents results of a  curiosity-driven research, which has been achieved only through the personal resources of the authors.

\section*{References}
\bibliographystyle{abrv-unsrturl}

\bibliography{Bibiliography_COVID19}

\bigskip 
\appendix
\section{Effective confirmed cases}
\label{apx: A}
\begin{figure}[h!]
    \center
    \includegraphics[width=\columnwidth]{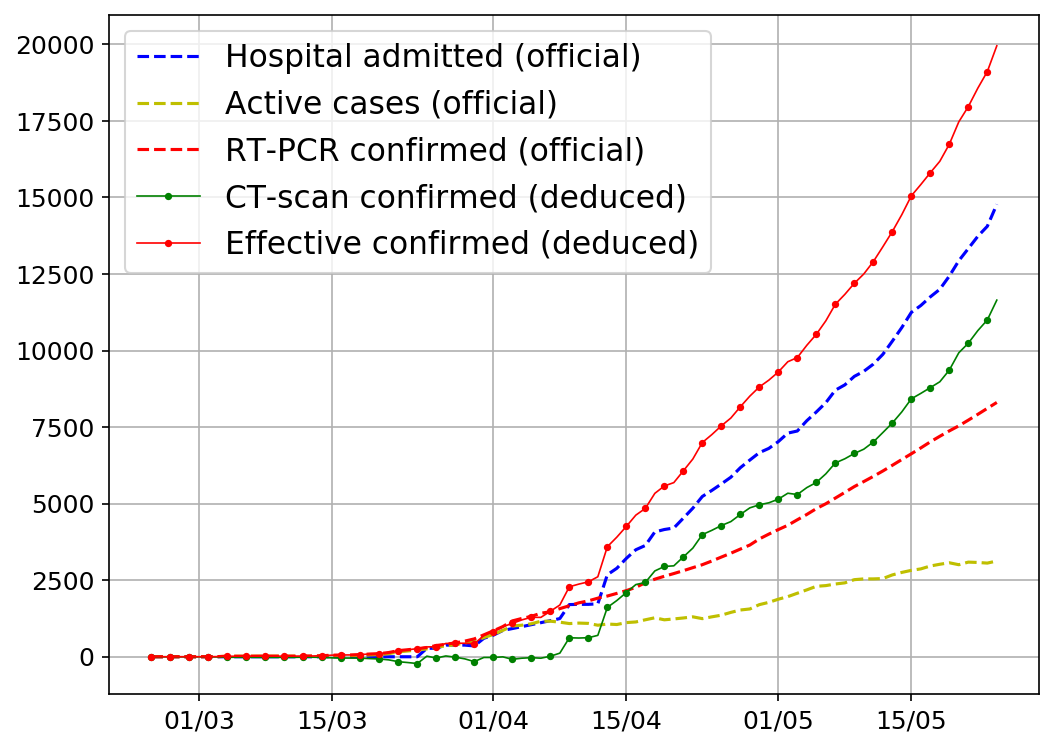}
    \caption{Estimated effective confirmed cases curve (dotted red) compared to official RT-PCR confirmed cases curve (dashed red) between Feb. 25th and May 24th, 2020. The number of CT-scan confirmed cases (green) for a given date is deduced by subtracting the number of official active cases (dashed yellow) from the number of hospital admitted patients (dashed blue) for that date.}
    \label{fig: A_vs_H}
\end{figure}

On Fig. \ref{fig: A_vs_H}, we expose the official number of hospital admitted patients due to COVID-19 in Algeria (blue dashed line) and the official number of active cases (yellow dashed line) computed by subtracting the official numbers for recoveries an deaths from the official RT-PCR-confirmed cases. Notice that before April 6th the two curves are perfectly superimposed. The number of hospital admitted patients start to significantly increase compared to the official number of active cases after April 6th. Indeed, since that time, the CT-scan diagnostic of COVID-19 have been adopted by Algerian health authorities in addition to the RT-PCR tests. Hence, we believe the difference between the two curves represent the number of CT-scan confirmed cases (green line). The deduced number of CT-scan confirmed cases is added to the official RT-PCR-confirmed cases to obtain the effective number of confirmed cases (red dotted line). Notice that the active cases curve displays a plateau on the beginning of April which is in our opinion due to the fact that RT-PCR testing capacity’s maximum limit has been reached. This curve started to increase again on the second half of April after many hospitals and biology research entities started performing RT-PCR tests \cite{INSP-1}.

\end{document}